\documentclass[a4paper]{jpconf}

\usepackage{amsmath,amssymb,amsthm}
\usepackage{graphicx}
\usepackage{longtable}
\usepackage{float}
\usepackage{textcomp}
\usepackage{slashed}

\newcommand{\forests}{\mathfrak{F}}
\newcommand{\infragr}{\mathfrak{I}'}

\begin{document}
\title{A Flexible Divergence Elimination Method for Calculating Lepton Magnetic Moments in Quantum Electrodynamics}
\author{Sergey Volkov}
\address{$^1$ Skobeltsyn Institute of Nuclear Physics of Lomonosov Moscow State University, Leninskie gory 1(2), GSP-1, 119991 Moscow, Russia}
\address{$^2$ Dzhelepov Laboratory of Nuclear Problems of Joint Institute for Nuclear Research, Joliot-Curie 6, 141980 Dubna, Moscow region, Russia}
\ead{volkoff\underline{ }sergey@mail.ru, sergey.volkov.1811@gmail.com}
\begin{abstract}
A precise calculation of the lepton anomalous magnetic moments (AMM) requires an evaluation of the quantum electrodynamics (QED) Feynman diagrams up to five independent loops. The complicated structure of ultraviolet (UV), infrared (IR) and mixed divergences in the corresponding integrals makes it difficult to calculate these high-order contributions in reasonable computer time frame. 

We demonstrate a method that eliminates all divergences point by point in Feynman parametric space (before integration) and possesses a flexibility that can be used for improving the precision of the numerical integration. This flexibility is especially actual for calculating the contributions of the Feynman diagrams with electron loops to the muon AMM. 3-loop and 4-loop numerical test results are provided.

The subtraction procedure is based on a forest formula with linear operators applied to the Feynman amplitudes of UV divergent subdiagrams. It is similar to BPHZ; the difference is in the linear operators used and in the way of combining them. It is equivalent to the on-shell renormalization after summation over diagrams: no residual renormalization is required. 
\end{abstract}

\section{Introduction}

The electron and muon anomalous magnetic moments are measured with very high precision. The values
\begin{align*}
a_e[\text{expt.}]&=0.001\,159\,652\,180\,73(28), \\
a_{\mu}[\text{expt.}]&=0.001\,165\,920\,61(41)
\end{align*}
are given in ~\cite{experiment} and ~\cite{muon_experiment}\footnote{Actually, the value is the statistical average of the new obtained value and the old one ~\cite{muon_experiment_old}.}, respectively. The Standard Model theoretical predictions for $a_e$ use the following representation:
$$
a_e=a_e(\text{QED})+a_e(\text{hadronic})+a_e(\text{electroweak}),
$$
$$
a_e(\text{QED})=\sum_{n\geq 1} \left(\frac{\alpha}{\pi}\right)^n
a_e^{2n},
$$
$$
a_e^{2n}=A_1^{(2n)}+A_2^{(2n)}(m_e/m_{\mu})+A_2^{(2n)}(m_e/m_{\tau})+A_3^{(2n)}(m_e/m_{\mu},m_e/m_{\tau}),
$$
where $m_e,m_{\mu},m_{\tau}$ are the masses of the electron, muon, and tau-lepton, $\alpha$ is the fine-structure constant. An analogous formula can be written for $a_{\mu}$. Different parts of these expressions were calculated by different researchers\footnote{See a review in ~\cite{kinoshita_atoms} for $a_e$ (by 2019) and in ~\cite{muon_sm_review} for $a_{\mu}$.}. The current theoretical value of $a_e$ is in agreement with the experiment, but there is a discrepancy about $4.2\sigma$ in $a_{\mu}$. The term $A_1^{(10)}$ is not double-checked yet, and a significant error in this term can affect the relationship between the theory and experiment for $a_e$. In 2019 the author recalculated the contribution of all Feynman diagrams without lepton loops to $A_1^{(10)}$ ~\cite{volkov_5loops_prd} and discovered a discrepancy about $4.8\sigma$ with the previously known value~\cite{kinoshita_atoms}. See details about the relationship between these calculations and different measurements of $\alpha$ in ~\cite{volkov_ffk2021}. The muon value $a_{\mu}$ is more sensitive to the hadronic corrections, but the value of $A_2^{(10)}(m_{\mu}/m_e)$ presented in ~\cite{kinoshita_muon} is not double-checked yet, and a significant error in it still can be sensitive in experiments\footnote{The muon QED coefficients suffer from powered large logarithms of $m_{\mu}/m_e$; this is why the QED calculations are actual for $a_{\mu}$ as well; for example, $a_e^{10}\approx 6.733$, $a_{\mu}^{10}\approx 750.86$.}.

\section{Old and new (flexible) method}

The on-shell renormalization removes all divergences in the QED terms of the lepton magnetic moments. However, the renormalization applied in place leaves individual diagrams IR divergent starting with two loops ~\cite{analyt2_p}.

There are several methods of removing divergences before integration suitable for calculation of the QED contributions to the lepton magnetic moments ~\cite{levinewright,carrollyao,kinoshita_infrared,kinoshita_automated}. Those methods remove divergences point by point in Feynman parametric space.

Our subtraction procedure presented in 2016 ~\cite{volkov_2015} will be referred to as the old method. That method has some advantages and was used in the above mentioned 5-loop calculation ~\cite{volkov_5loops_prd}. It uses the Feynman parameters as well. If by $N_l$ and $N_{\gamma}$ we denote the number of external lepton and photon lines of a diagram, in QED we have four types of UV divergent subdiagrams\footnote{Only one-particle irreducible subdiagrams are considered.}: vertex-like ($N_l=2,N_{\gamma}=1$), electron self-energy ($N_l=2,N_{\gamma}=0$), photon self-energy ($N_l=0,N_{\gamma}=2$), photon-photon scattering ($N_l=0,N_{\gamma}=4$). The old method uses the following three linear operators applied to the Feynman amplitudes of UV divergent subdiagrams:
\begin{itemize}
\item $A$ is the anomalous magnetic moment projector (multiplied by the Dirac matrix $\gamma_{\mu}$) applied to vertex-like Feynman amplitudes;
\item $U$ is an intermediate operator defined as
$$
U\Gamma_{\mu}(p,q)=a(m^2)\gamma_{\mu},\quad U\Sigma(p)=r(m^2)+s(m^2)\slashed p,
$$
where $\Gamma_{\mu}(p,q)$ and $\Sigma(p)$ are vertex-like\footnote{$p-\frac{q}{2}$, $p+\frac{q}{2}$ are the ingoing and outgoing lepton momenta, $q$ is the photon momentum.} and lepton self-energy Feynman amplitudes,
\begin{equation}\label{eq_gamma_sigma}
\Gamma_{\mu}(p,0)=a(p^2)\gamma_{\mu}+b(p^2)p_{\mu}+c(p^2)\slashed{p}p_{\mu}+d(p^2)(\slashed{p}\gamma_{\mu}-\gamma_{\mu}\slashed{p}),\quad \Sigma(p)=r(p^2)+s(p^2)\slashed{p},
\end{equation}
$m$ -- is the mass of the external lepton for this subdiagram; we use the metric tensor $g_{\mu\nu}$ such that $g_{00}=1$, $g_{11}=g_{22}=g_{33}=-1$. For the other types of UV divergent subdiagrams $U$ is defined as the Taylor expansion around zero momenta up to the needed order (for UV divergence elimination);
\item $L$ is the on-shell renormalization projector applied to vertex-like Feynman amplitudes:
$$
L\Gamma_{\mu}(p,q)=[a(m^2)+b(m^2)m+c(m^2)m^2]\gamma_{\mu}.
$$
\end{itemize}
See the complete definitions in ~\cite{volkov_2015}. The old method is a fixed subtraction procedure with the fixed operators $A$, $L$, $U$\footnote{See ~\cite{volkov_2015} for the formulation. See also an analysis of the flaws and a comparison with the new method in ~\cite{volkov_ffk2021}.}.

The new method was first formulated in ~\cite{volkov_ffk2021}. It uses four operators $U_0,U_1,U_2,U_3$ instead of $U$; the operators $U_1,U_2,U_3$ are not fixed, but are given by requirements; see a preliminary (not complete) list of requirements in ~\cite{volkov_ffk2021}. Also, these operators may be different for different types of subdiagram external leptons. The divergence subtraction procedure is an object of a very accurate tuning, and it is difficult to describe all possible freedom in the procedure, because it must remove all divergences in each individual diagram (including mixed UV-IR divergences\footnote{See an example of mixed divergences in ~\cite{adkins}.}) and simultaneously it must be equivalent to the on-shell renormalization\footnote{Thus, the combinatorics of the procedure must contain an evidence that the on-shell renormalization removes all divergences.}.

Let $G$ be a Feynman diagram contributing to the lepton AMM. By $\infragr[G]$ we denote the set of all vertex-like subgraphs of $G$ (including $G$) lying on the main path\footnote{The main path of $G$ is the lepton path connecting its external lepton lines.} of $G$ and having the external photon of $G$. By $\forests[G]$ we denote the set of all forests\footnote{A forest is a set of subdiagrams such that each of them are nested or non-intersecting.} of UV-divergent subdiagrams of $G$ containing $G$. The expression for subtraction and extraction of AMM is
$$
\sum_{\substack{F=\{G_1,\ldots,G_n\}\in\forests[G] \\ G'\in \infragr[G] \cap F }} (-1)^{n-1} M^{G'}_{G_1} M^{G'}_{G_2} \ldots M^{G'}_{G_n}.
$$
Here $M^{G'}_{G''}$ equals $A_{G''}$, if $G''=G'$; $L_{G''}-(U_1)_{G''}$, if $G''=G\neq G'$; $L_{G''}$, if $G'\subset G''\subset G$;  $(U_0)_{G''}$, if $G''$ is a photon self-energy or a photon-photon subgraph. In the remaining cases it equals $(U_2)_{G''}$, if $G''$ lies on a lepton loop; $(U_w)_{G''}$, if $G''\in\infragr[G]$ and $G''\subset G'$; $(U_1)_{G''}$ in the other cases; here $w=3$, if $G$ has its external photon on a lepton loop, $w=1$ otherwise. We should take the coefficient before $\gamma_{\mu}$ after applying the expression. $U_0$ is defined as the Taylor expansion around zero momenta up to the needed order ($2$ for photon self-energy, $0$ for photon-photon scattering); we can also use other known treatments of photon self-energy subdiagrams.

\begin{figure}[H]
	\begin{center}
		\includegraphics[width=60mm]{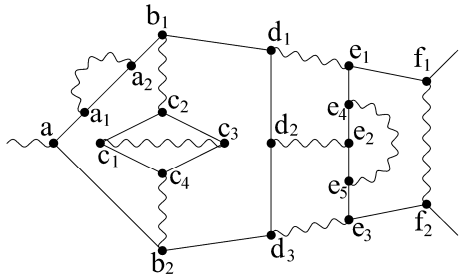}
		\caption{\label{fig_example}An example of a Feynman diagram contributing to AMM.}
	\end{center}
	
	%\vspace{-5mm}
\end{figure}

For example, the diagram $G$ from Figure \ref{fig_example} gives the expression
\begin{gather*}
\left[A_G \left( 1-(U_3)_{G_e} \right) - \left( L_G - (U_1)_G \right) A_{G_e} \right] \times \left( 1-(U_2)_{G_c} \right) \times \left(1 - (U_1)_{e_2 e_4 e_5} \right) \\
\times \left(1-(U_0)_{G_d}\right)\times \left(1-(U_0)_{c_1c_2c_3c_4} \right) \times \left( 1 - (U_2)_{c_1 c_2 c_3} - (U_2)_{c_1 c_3 c_4} \right) \times \left( 1- (U_2)_{a_1a_2} \right),
\end{gather*}
where $G_e=aa_1a_2b_1b_2c_1c_2c_3c_4d_1d_2d_3e_1e_2e_3e_4e_5$, $G_d=\{aa_1a_2b_1b_2c_1c_2c_3c_4d_1d_2d_3\}$, $G_c=aa_1a_2b_1b_2c_1c_2c_3c_4$, $\infragr[G]=\{G_e,G\}$, subdiagrams are denoted by enumeration of their internal vertices; we should expand brackets for obtaining the forest formula. The expression for this example does not depend on the type of leptons on lepton loops (however, each of the operators $U_1,U_2,U_3$ can be defined differently for different types of leptons).

The algorithm of calculating a diagram set contribution to the lepton AMM looks as follows:
\begin{itemize}
\item For the given Feynman diagram introduce the Schwinger parametric propagators
$$
(\slashed{p}+m)\exp(i\alpha_j(p^2-m^2+i\varepsilon)),\quad ig_{\mu\nu}\exp(i\alpha_j(p^2+i\varepsilon))
$$
for the lepton and photon lines, respectively, where $\alpha_j$ is the parameter corresponding to $j$-th internal line, $\varepsilon>0$ is an infrared regulator.
\item Write the expression with linear operators (defined above) and obtain the result of applying this expression as a function of $\alpha_1,\ldots,\alpha_n,\varepsilon$ (by using known explicit formulas for integrals of multidimensional gaussian functions multiplied by polynomials ignoring the integral divergence; in the case of nested subdiagrams with operators, we can replace the subdiagram Feynman amplitudes with their images under the corresponding operators sequentially from smaller to larger subdiagrams).
\item Obtain the result of applying the expression as a function of the Feynman parameters $z_1,\ldots,z_n$ and $\varepsilon$ (taking into account the known relationship between the Schwinger and Feynman parameters\footnote{See, for example, ~\cite{bogoliubov_shirkov_quantum_fields}.}, we should only make the change of variables $\alpha_j=\lambda z_j$ and perform the intergration over $\lambda$ (from $0$ to $+\infty$) analytically).
\item Take the limit $\varepsilon\rightarrow +0$ (analytically).
\item Integrate over $z_1,\ldots,z_n$ numerically ($z_j\geq 0$, $z_1+\ldots+z_n=1$).
\item Calculate the sum over all needed Feynman diagrams.
\end{itemize}
The details of the algorithm are described in ~\cite{volkov_2015}.

\section{Tests and discussion}

We will use the operators $U_j$, $j=1,2,3$ of the form
$$
U_j\Gamma(p,q)=a(M^2)\gamma_{\mu},\quad U_j\Sigma(p)=r(m^2)+s(m^2)m+s(M^2)(\slashed{p}-m),
$$
where $M^2$ is some subtraction point, (\ref{eq_gamma_sigma}) is satisfied, $m$ is the external lepton mass in the subdiagram to which the operator is applied. $U_j$ preserves the Ward identity and extracts the mass part of $\Sigma(p)$ completely; thus, we can consider the usage of these operators for our purposes ~\cite{volkov_ffk2021}. 

Since the most effective way of numerical integrating functions of many variables is the Monte Carlo method, we can compare different choices of $M^2$ with respect to the Monte Carlo convergence speed. If some value is obtained as the integral $\int f(z) dz$, the value $\int |f(z)|dz$ can be used as an estimation of the ``badness'' of $f(z)$ for Monte Carlo integration. Also, there is an observation that the contributions of individual diagrams are often ``oscillating'' as well as the Feynman parametric integrands~\cite{cvitanovic_gauge,volkov_5loops_prd} regardless of the subtraction method used, and the oscillations are cancelled after summation over all diagrams in a gauge-invariant set. The value $\int |f(z)|dz$ (and its sum over a set) can be used as an estimation of these oscillations.

\begin{figure}[H]
	\begin{center}
		\includegraphics[width=70mm]{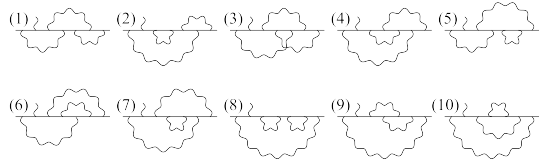}
		\caption{\label{fig_3loops_gauge_set}A 3-loop gauge-invariant set.}
	\end{center}
	
	%\vspace{-5mm}
\end{figure}

Table \ref{table_3loops_set} contains the contributions of the 3-loop diagrams from Figure \ref{fig_3loops_gauge_set} to $A_1^{(6)}$ and the corresponding estimations for Monte Carlo integration obtained with $M^2=m^2$ and $M^2=-m^2$. The final results are in good agreement with each other\footnote{The difference in the precisions is due to the difference in the computer time and hardware used.} and with the known analytical result $0.53336$\footnote{See a detailed three-loop comparison with analytical results in ~\cite{volkov_ffk2021}.}. A general observation is that space-like subtraction points $M^2<0$ produce less oscillations than time-like ones.

\begin{table}[H]
\centering\tiny %\addtolength\tabcolsep{-0.5mm}
\caption{\label{table_3loops_set}\footnotesize Contributions of the diagrams from Figure \ref{fig_3loops_gauge_set} to $A_1^{(6)}$ obtained with different subtraction points and the corresponding integrals of the absolute values.}	%\texttt{New 2}
%\medskip
\begin{tabular}{ccccc}
\hline \hline \\[-1.8mm] Diagram & $M^2=m^2$, value & $M^2=-m^2$, value & $M^2=m^2$, $\int |f(z)|dz$ & $M^2=-m^2$, $\int |f(z)|dz$ \\ \hline \\[-1.8mm]
1 & $-1.68013(35)$ & $-0.97296(81)$ & $2.105$ & $1.394$ \\
2 & $-0.09677(49)$ & $-0.6103(18)$ & $1.667$ & $1.311$ \\
3 & $0.21489(27)$ & $-0.48545(34)$ & $0.819$ & $0.86$ \\
4 & $0.14480(27)$ & $0.8527(27)$ & $1.111$ & $1.01$ \\
5 & $0.83283(40)$ & $1.1325(19)$ & $1.33$ & $1.251$\\
6 & $-0.02875(23)$ & $0.2158(25)$ & $0.666$ & $0.496$  \\
7 & $0.80395(54)$ & $1.10398(60)$ & $1.31$ & $1.173$ \\
8 & $-2.12293(27)$ & $-1.2838(17)$ & $2.123$ & $1.318$ \\
9 & $2.52480(29)$ & $1.1537(41)$ & $2.587$ & $1.413$ \\
10 & $-0.05880(20)$ & $-0.5792(27)$ & $1.084$ & $1.269$ \\
$\sum$ & $0.5339(11)$ & $0.5272(70)$ & $14.8$ & $11.49$ 
\\ \hline \hline
\end{tabular}
\end{table}

The contributions of the 4-loop classes IV(b) and IV(c) with an electron loop ~\cite{kinoshita_muon} to the muon term $A_2^{(8)}(m_{\mu}/m_e)$\footnote{For these tests we use the fixed values $m_{\mu}=105.6583745\,\text{MeV}$, $m_e=0.51099895\,\text{MeV}$ and do not take into account their uncertainty.} are evaluated in Tables \ref{table_IV_b} and \ref{table_IV_c}. The first table contains the evaluation with different subtraction points $M^2=(m_e)^2$ and $M^2=-m_e m_{\mu}$. The results are in good agreement with each other\footnote{However, the precision is not so high because of a relatively small computer time and an unadjusted Monte Carlo integration algorithm.} and with the known values $-0.4170(37)$ and $-0.38(8)$ from ~\cite{kinoshita_muon} and ~\cite{smirnov_mu_with_e}, respectively. The second choice of the subtraction point produces less oscillations, this can be observed visually. The value $M^2=-m_{\mu}m_e$ is almost optimal: the selections $M^2=-(m_{\mu})^2$, $M^2=-m_{\mu}m_e$, $M^2=-(m_e)^2$, $M^2=(m_e)^2$ give the sums of $\int |f(z)|dz$ equal $1374$, $767$, $1113$, $1510$, respectively.

\begin{figure}[H]
	\begin{center}
		\includegraphics[width=40mm]{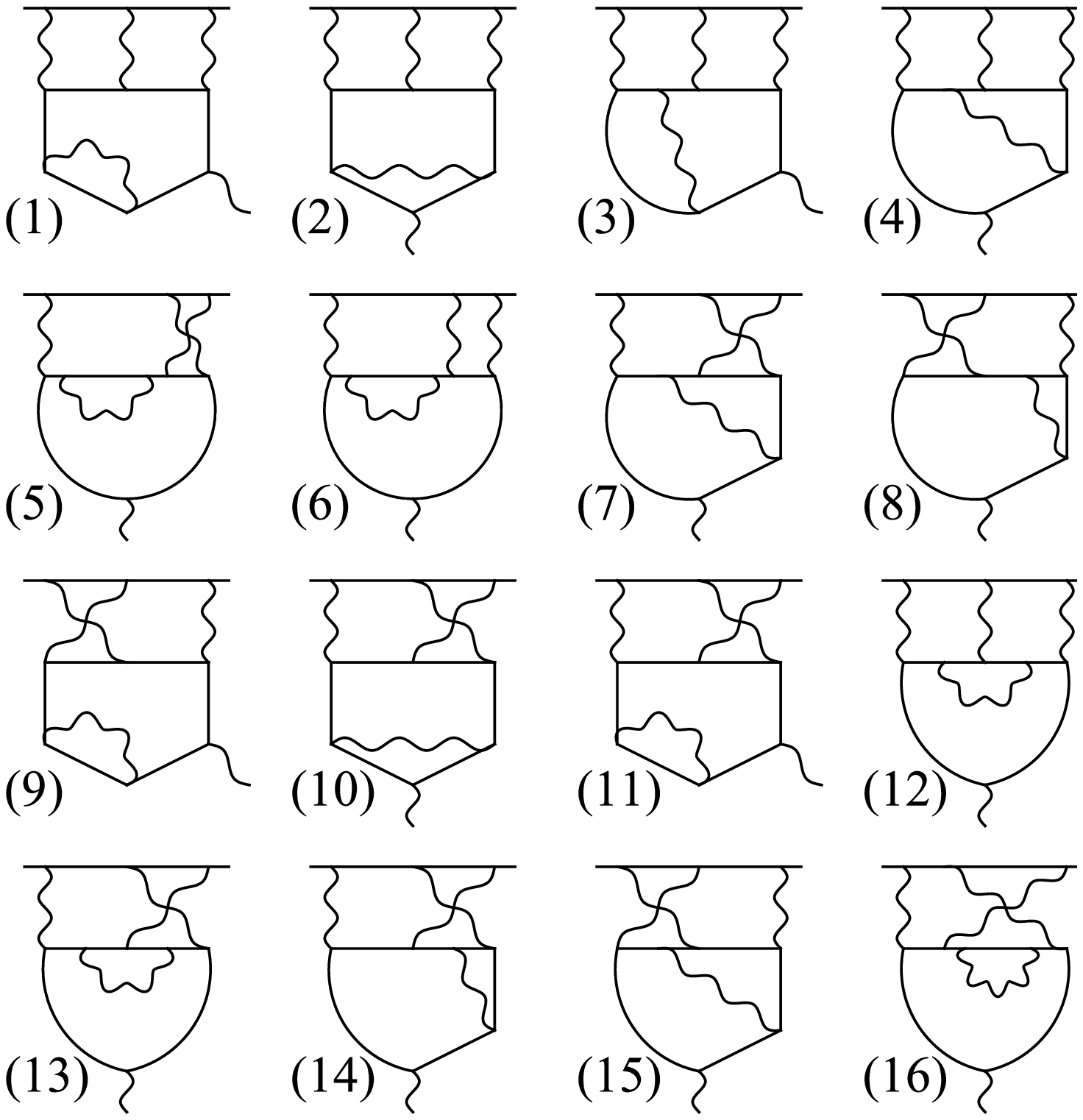}
		\ \ \ \ \ \ \ \ 
		\includegraphics[width=35mm]{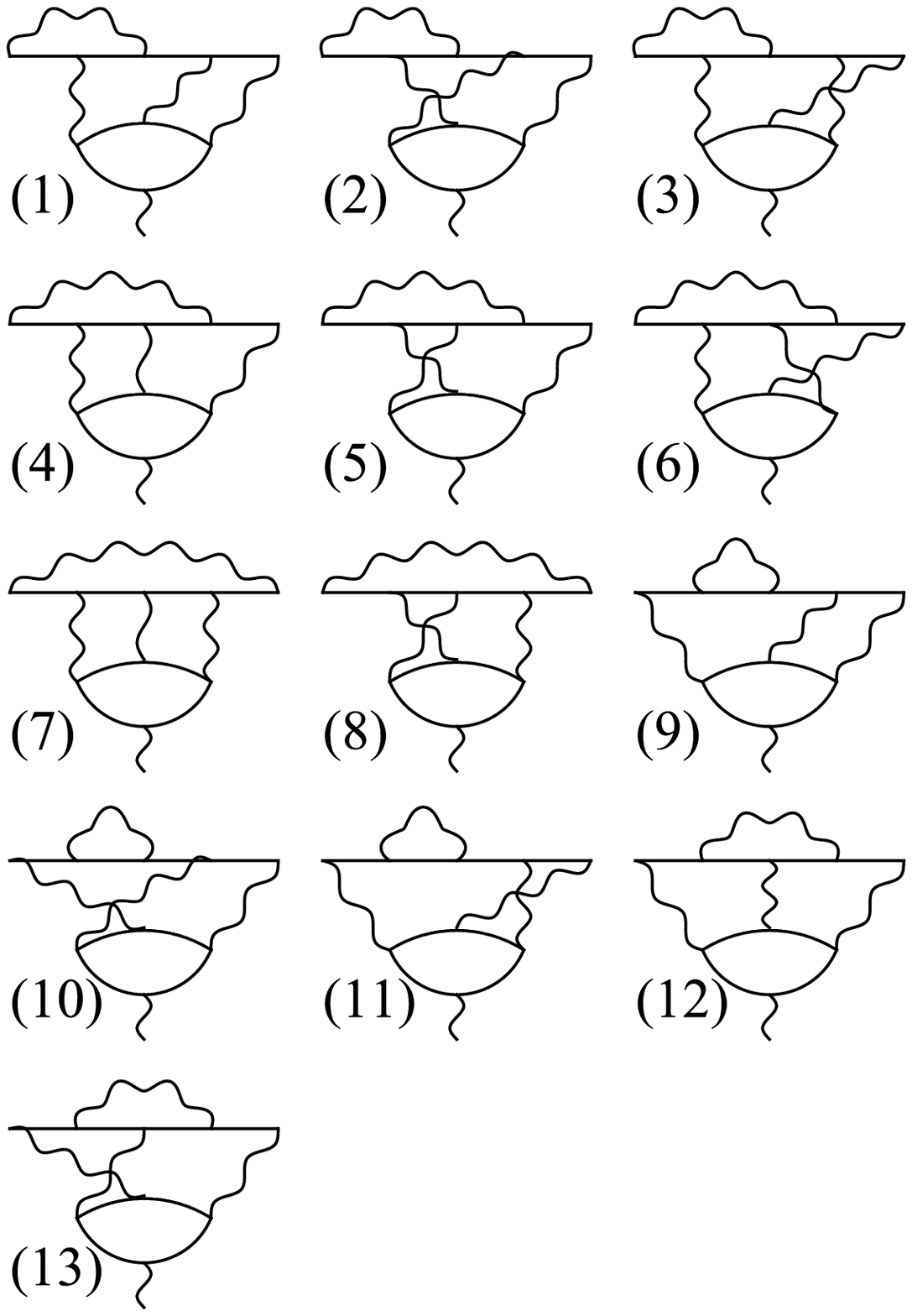}
		\vspace{-3mm}
		\caption{\label{fig_IV_b_c}4-loop gauge-invariant classes IV(b) (left) and IV(c) (right) from ~\cite{kinoshita_muon}.}
	\end{center}
	
	%\vspace{-5mm}
\end{figure}

\begin{table}[H]
\centering\tiny %\addtolength\tabcolsep{-0.5mm}
\caption{\label{table_IV_b}\footnotesize Contributions of the diagrams from IV(b) (Figure \ref{fig_IV_b_c}, left) to $A_2^{(8)}(m_{\mu}/m_e)$ obtained with different subtraction points.}	%\texttt{New 2}
%\medskip
\begin{tabular}{cccccc}
\hline \hline \\[-1.8mm] \textnumero & $M^2=(m_e)^2$ & $M^2=-m_e m_{\mu}$& \textnumero & $M^2=(m_e)^2$ & $M^2=-m_e m_{\mu}$ \\ \hline \\[-1.8mm]
1 & $32.712(73)$ & $-2.753(13)$ & 10 & $4.92(12)$ & $20.465(20)$ \\
2 & $-7.118(63)$ & $10.580(11)$ & 11 & $20.817(95)$ & $5.318(18)$ \\
3 & $-27.870(89)$ & $7.688(17)$ & 12 & $-7.745(98)$ & $9.902(17)$ \\
4 & $-23.811(15)$ & $-23.791(11)$ & 13 & $-11.21(13)$ & $4.371(27)$ \\
5 & $26.517(93)$ & $11.071(18)$ & 14 & $-19.36(11)$ & $-3.791(22)$ \\
6 & $39.12(10)$ & $3.642(19)$ & 15 & $-26.433(25)$ & $-26.417(11)$ \\
7 & $-23.529(21)$ & $-23.522(13)$ & 16 & $25.86(10)$ & $10.431(22)$ \\
8 & $-20.30(13)$ & $-4.753(22)$ & $\sum$ & $-0.73(37)$ & $-0.340(72)$ \\
9 & $16.70(11)$ & $1.221(16)$ & \ & \ & \ 
\\ \hline \hline
\end{tabular}
\end{table}

Table \ref{table_IV_c} demonstrates an evaluation of the contribution of IV(c) to $A_2^{(8)}(m_{\mu}/m_e)$ with two different combinations of the linear operators: $U_1=U_2=U,U_3=L$; $U_1=U_2=U_3=U$. The combinations give the same values for all diagrams except 7 and 8: they are the only diagrams with different expressions. However, we have a good agreement after summation between these calculations and with the known values $2.9072(44)$ from ~\cite{kinoshita_muon} and $2.94(30)$ from ~\cite{smirnov_mu_with_e}.

The author thanks Lidia Kalinovskaya, Gudrun Heinrich, Savely Karshenboim, Andrey Arbuzov, Andrey Kataev for the important assistance. Also, the author thanks the Laboratory of Information Technologies of JINR (Dubna, Russia) for providing an access to its computational resources and additionally the organizers of the conference ACAT-2021 for providing a possibility to make an online presentation.

\begin{table}[H]
\centering\tiny %\addtolength\tabcolsep{-0.5mm}
\caption{\label{table_IV_c}\footnotesize Contributions of the diagrams from IV(c) (Figure \ref{fig_IV_b_c}, right) to $A_2^{(8)}(m_{\mu}/m_e)$ obtained with different selections of $U_3$ (we use $U_1=U_2=U$).}	%\texttt{New 2}
%\medskip
\begin{tabular}{cccccc}
\hline \hline \\[-1.8mm] \textnumero & $U_3=L$ & $U_3=U$ & \textnumero & $U_3=L$ & $U_3=U$ \\ \hline \\[-1.8mm]
1 & $-118.179(22)$ & $-118.167(22)$ & 8 & $20.775(32)$ & $4.920(30)$ \\
2 & $91.841(21)$ & $91.778(21)$ & 9 & $100.475(24)$ & $100.514(25)$ \\
3 & $-80.820(20)$ & $-80.839(21)$ & 10 & $-76.924(32)$ & $-76.923(33)$ \\
4 & $2.620(22)$ & $2.612(25)$ & 11 & $89.140(31)$ & $89.186(35)$ \\
5 & $-74.730(26)$ & $-74.694(29)$ & 12 & $-26.766(12)$ & $-26.763(12)$ \\
6 & $37.016(22)$ & $37.015(24)$ & 13 & $-10.543(21)$ & $-10.550(20)$ \\
7 & $48.934(19)$ & $64.810(19)$ & $\sum$ & $2.840(86)$ & $2.899(90)$  
\\ \hline \hline
\end{tabular}
\end{table}

\end{document}